\documentclass[twocolumn,prd,nofootinbib,showpacs]{revtex4}
\usepackage{graphicx}
\usepackage{color}
\usepackage{amsmath,amssymb,amsfonts,dsfont}
\usepackage{multirow}
\usepackage{sistyle}
\usepackage{ulem}

\def\26{{$\sigma_{26}$}}

\newcommand{\be}{\begin{equation}}
\newcommand{\ee}{\end{equation}}
\newcommand{\bea}{\begin{eqnarray}}
\newcommand{\eea}{\end{eqnarray}}
\newcommand{\sigmav}{$\langle\sigma v\rangle$ }

\newcommand{\lsim}{\mathrel{\mathop{\kern 0pt \rlap
  {\raise.2ex\hbox{$<$}}}
  \lower.9ex\hbox{\kern-.190em $\sim$}}}
\newcommand{\gsim}{\mathrel{\mathop{\kern 0pt \rlap
  {\raise.2ex\hbox{$>$}}}
  \lower.9ex\hbox{\kern-.190em $\sim$}}}

\begin{document}
\title{Constraining  dark matter annihilation with the isotropic $\gamma$-ray background:\\
 updated limits and future potential}
\author{Torsten Bringmann}\email{Torsten.Bringmann@fys.uio.no}
\affiliation{ II. Institute for Theoretical Physics, University of Hamburg, Luruper Chaussee 149, 22761 Hamburg, Germany\\
Department of Physics, University of Oslo, Box 1048 NO-0316 Oslo, Norway}
\author{Francesca Calore}\email{francesca.calore@desy.de}
\affiliation{ II. Institute for Theoretical Physics, University of Hamburg, Luruper Chaussee 149, 22761 Hamburg, Germany\\
GRAPPA Institute, University of Amsterdam, Science Park 904, 1090 GL Amsterdam, The Netherlands}
\author{Mattia Di Mauro}\email{mattia.dimauro@to.infn.it}
\affiliation{Dipartimento di Fisica, Universit\`a di Torino
and INFN, Sezione di Torino, Via P. Giuria 1, 10125 Torino, Italy}
\author{Fiorenza Donato}\email{donato@to.infn.it}
\affiliation{Dipartimento di Fisica, Universit\`a di Torino
and INFN, Sezione di Torino, Via P. Giuria 1, 10125 Torino, Italy}

\begin{abstract}
The nature of the Isotropic $\gamma$-ray Background (IGRB) measured by the Large Area Telescope (LAT) on the 
 Fermi $\gamma$-ray space Telescope ({\it Fermi})  remains partially unexplained. 
Non-negligible contributions may originate from extragalactic populations of unresolved sources such as 
  blazars,  star-forming galaxies or galactic  milli-second pulsars.
A recent prediction of the diffuse $\gamma$-ray emission from Active Galactic Nuclei (AGN) 
with a large viewing angle with respect to the line-of-sight (l.o.s.)  has demonstrated 
that this faint but numerous population is also expected to  contribute significantly to the total
IGRB intensity. 
A more exotic contribution to the IGRB invokes the pair annihilation of dark matter (DM) weakly interacting 
massive particles (WIMPs) into $\gamma$-rays. 
In this work, we evaluate the room left for galactic DM at high latitudes  ($>10^\circ$) 
by including photons from both prompt emission and inverse Compton scattering, emphasizing the impact of  the newly discovered contribution from misaligned AGN (MAGN) for 
such an analysis.  Summing up all significant galactic and extragalactic components of the IGRB, we find that an improved understanding of the associated astrophysical uncertainties 
is  still mandatory to put stringent bounds on thermally produced DM. On the other hand, we also demonstrate that the IGRB has the potential to be one of the most competitive {\it future} 
ways to test the DM WIMP hypothesis, once the present uncertainties are even slightly reduced. In fact, if MAGN contribute even at 90\% of the maximal level consistent with our current 
understanding, thermally produced WIMPs would be severely constrained as DM candidates for masses up to several TeV.

  \end{abstract}
\maketitle

\section{Introduction}
The nature of dark matter (DM) remains one of the most intriguing 
mysteries in fundamental physics and cosmology. A viable investigation technique 
consists in the search for its stable annihilation products in the halo of galaxies, and in particular 
in the Milky Way. Making the common hypothesis for DM to consist of weakly interacting 
massive particles (WIMPs),  one of the most promising indirect detection 
channels is through  annihilation into $\gamma$-rays (see Ref.~\cite{Bringmann:2012ez} for a recent review). 
Unprecedented $\gamma$-ray observations of diffuse emission and point sources by the {\it Fermi}-LAT \cite{Atwood:2009ez}
have stimulated the search for exotic components from DM annihilation 
in the Milky Way, in extragalactic nearby objects,  as well as
in cosmological structures \cite{2012ApJ...761...91A,2010ApJ...712..147A,2010JCAP...05..025A,Abdo:2010dk, 2011PhRvL.107x1302A}. 
The incontrovertible DM signature would be a $\gamma$-ray monochromatic line at 
energies around the WIMP mass.
A recent analysis of {\it Fermi}-LAT data of the galactic center region 
has unveiled the feature of a narrow line close to 130 GeV
 \cite{2012JCAP...07..054B,2012JCAP...08..007W,Tempel:2012ey,2013JCAP...01..029F}, 
with an angular 
distribution consistent
with the expectation for WIMP DM \cite{Bringmann:2012ez} and a statistical significance of at least 3$\sigma$ after trials 
factor correction (in fact, more than 5$\sigma$ when taking into account both spectral {\it and} spatial information of the 
incoming events \cite{2013JCAP...01..029F}). While confirming this excess, the  
{\it Fermi}-LAT Collaboration finds that its significance is substantially reduced to a mere 1.6$\sigma$
(albeit in somewhat different target regions)
when considering a larger set of re-processed data and a 2D modeling of the energy dispersion 
\cite{Bloom:2013mwa,Fermi-LAT:2013uma}. 
The interpretation in terms of  annihilating DM would in any case require
 the loop-suppressed direct annihilation into photons to be larger than typically expected for thermally produced DM, 
 a fact which has already inspired considerable model-building efforts 
 (see, e.g.,  the compilation in Refs.~\cite{Bringmann:2012ez,Weniger:2013tza} for an overview). 
 Only a  larger exposure will allow to finally settle possible issues concerning the statistical 
 robustness and potential instrumental effects of the line feature; with the proposed new observation strategy 
 for {\it Fermi}, recently positively reviewed by the collaboration, this could be achievable until the end of 2014 \cite{Weniger:2013tza}.

A faint  $\gamma$-ray emission at high latitudes ($|b|>10^\circ$) has been measured by {\it Fermi}-LAT \cite{IDGRB},
 showing a high degree of isotropy \cite{Ackermann:2012uf} and an energy spectrum compatible with 
a power-law. Though  its very nature is still unexplained, it is   
reasonable to believe it to be the superposition of (mainly) extragalactic unresolved point-sources and, to a lesser extent, 
of astrophysical diffuse processes  (see Ref.~\cite{2012PhRvD..85b3004C} and Section \ref{sect:IGRB} for a detailed discussion). 
A  possible contribution from DM annihilation, both from galactic and cosmological distances,
can be hidden in this isotropic $\gamma$-ray background (IGRB) data as well 
\cite{Bergstrom:2001jj,Ullio:2002pj,Taylor:2002zd,Papucci:2009gd,Cirelli:2009dv,Baxter:2010fr,Abazajian:2010zb,Blanchet:2012vq,2012PhRvD..85b3004C}.
\footnote{
 The 130\,GeV feature described above does not show up in the IGRB  -- as  expected for the annihilation rate of 
 $(\sigma v)_{\gamma\gamma}\sim 10^{-27}{\rm cm}^3{\rm s}^{-1}$ inferred from the galactic center observation  \cite{Abdo:2010dk} 
 (unless one adopts very optimistic models for the evolution and distribution of DM subhalos \cite{Zavala:2009zr}). }
Previous upper limits on the DM annihilation cross section from the high latitudes 
$\gamma$-ray emission were derived e.g. in  \cite{2012PhRvD..85b3004C},  by subtracting from the IGRB \cite{IDGRB} 
the minimum estimated fluxes for the most significant unresolved source populations. 
\\
The faint but numerous population of Active Galactic Nuclei (AGN) whose jet axes have a large viewing angle 
with respect to the line-of-sight (l.o.s.) has recently been shown \cite{RG,moriond,2011ApJ...733...66I} 
to  contribute between 10\% and 100\% to the IGRB  intensity,  depending on the astrophysical uncertainties involved. 
In the present paper, we extend and update the analysis in Ref.~\cite{2012PhRvD..85b3004C} by 
 taking into account those new results on the isotropic diffuse flux predicted for misaligned AGN (MAGN) \cite{RG,moriond}. 
We find that the annihilation intensity of DM in the galactic halo may be severely constrained by the inclusion
of this new diffuse astrophysical background, depending on the WIMP mass and on the annihilation channel considered.

In this context, it is worth noting that classical DM indirect detection targets like the galactic center or dwarf spheroidal galaxies, which currently 
result in the best available limits on DM annihilation 
\cite{GeringerSameth:2011iw,2011PhRvL.107x1302A,Abramowski:2011hc,Hooper:2012sr,Tavakoli:2013zva,2012ApJ...761...91A}, 
will at some point inevitably face fundamental limitations in the sense that even greatly increased statistics will not allow to further 
improve those limits \cite{Bringmann:2012ez}.  For ground-based telescopes, this is likely the case already for anything beyond the planned 
Cherenkov Telescope Array \cite{Acharya:2013sxa}, while for future space-based missions there might be room for further improvement of the 
limits by up to one order of magnitude with respect to what can be expected by the end of the {\it Fermi} mission.
Once the MAGN-related astrophysical uncertainties are under better control, the IGRB may thus well turn out to be the new driving force in 
setting ever more stringent limits on the DM annihilation rate.

The paper is organized as follows. In Sect.\,\ref{sect:IGRB} we present the possible contributions to the IGRB and we discuss the relevance of 
unresolved extragalactic sources, notably MAGN. The expected galactic DM signal is described in Sect.\,\ref{sect:DM}. The DM spectrum is made up by the prompt 
$\gamma$-ray component and photons from inverse Compton scattering (ICS) of electrons and positrons produced by DM annihilation. 
By subtracting the astrophysical components from the high latitudes $\gamma$-ray emission, in Sect.\,\ref{sect:constraints} we derive 
 upper limits on the DM annihilation cross section for several annihilation channels and under various assumptions for the MAGN contribution.
We discuss the implications for DM indirect detection, as well as 
  the dependence on the other possible astrophysical contributions.
In Sect.\,\ref{sect:conclusions} we present our conclusions.

\section{Diffuse $\gamma$-ray emission from unresolved astrophysical sources }
\label{sect:IGRB}
\begin{figure}[t!]
\includegraphics[scale=0.65]{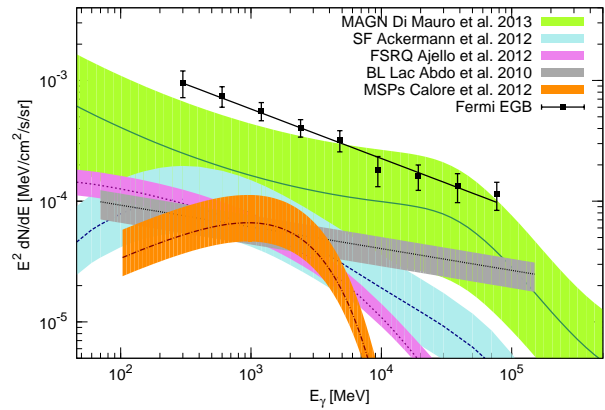}  
\caption{The diffuse $\gamma$-ray fluxes predicted for various unresolved point-source populations, along with 
the IGRB data from the {\it Fermi}-LAT Collaboration based on high galactic latitude ($|b|>10^{\circ}$) observations \cite{IDGRB}. 
The displayed contributions to the IGRB and corresponding uncertainty bands arise from: MAGN \cite{RG} (solid dark-green curve and light-green band), 
MW model for SF galaxies \cite{2012ApJ...755..164A} (dot-dashed red line and orange band), 
MSPs \cite{2012PhRvD..85b3004C} (dot-dashed red line and orange band), BL Lacs \cite{2010ApJBlazarFermi} (dotted dark-grey line and light-grey band), 
and FSRQs \cite{2012ApJ...751..108A} (purple dotted line and pink band).} 
\label{fig:fluxes}  
\end{figure}

The   IGRB found at high latitudes ($|b| > 10^{\circ}$) 
has been obtained by the {\it Fermi}-LAT Collaboration \cite{IDGRB} after
subtracting from the LAT data  the contributions from point-like sources, the galactic diffuse emission (which represents the major systematic 
uncertainty of that analysis), the background from CRs in the detector and solar photons. 
 The energy spectrum of the IGRB is consistent with a power-law having spectral index $\alpha = 2.41 \pm 0.05$ and integrated intensity
  $I(> 100$  {\rm MeV}$) = (1.03 \pm 0.17) \cdot 10^{-5}$ {\rm cm}$^{-2}$ {\rm s}$^{-1}$ {\rm sr}$^{-1}$.
This diffuse emission  likely  arises from the superposition of several contributions from unresolved  point-source populations, 
as well as from photons originating from truly diffuse processes (i.e~extended source contributions like from ultra-high energy cosmic rays,  
nearby galaxy clusters
or gravitationally induced shock waves during structure formation,  see Ref.~\cite{2012PhRvD..85b3004C} and references therein).  
Most of the photons in the IGRB are believed to be of extragalactic origin.  The truly diffuse processes are  subdominant and may
 be non-negligible (\% level) only at the high-energy end of the range explored by the {\it Fermi}-LAT.
 We have checked that the total UHECR diffuse emission predicted by \cite{2011PhLB..695...13B} 
 contributes to at most 0.3\%  of the measured IGRB. On the other side, the high energy  ($>$ 100 GeV) emission from galaxy clusters 
 predicted by \cite{2004APh....20..579G} is O(10$^{-2}$) of the IGRB at 100 GeV. 
 Given their very marginal role in the analysis undertaken here,   they will be neglected from now on. 
\\
{\it Misaligned AGN}. 
Recently, a thorough analysis of the $\gamma$-ray flux from unresolved MAGN 
has demonstrated that this numerous but faint population may contribute between 10\% and 100\% to
 the measured IGRB \cite{RG,moriond}. 
A jet displacement of about 14$^\circ$ marks the separation between blazars and non-blazar (namely misaligned) AGN. In the unification model 
\citep{urry1995},  radio galaxies  are those misaligned active galaxies whose jet points on average  at an angle $> 44^\circ$, while lower 
angles indicate typically steep spectrum radio quasars. 
The result in Refs.~\cite{RG,moriond} is based on the $\gamma$-ray data of 12 MAGN  by {\it Fermi}-LAT, 
and is based on  a correlation between their $\gamma$-ray  and  core radio luminosities. 
This correlation has been  tested against the MAGN  number count distribution, and validated by 
95\% flux upper limits on the {\it Fermi}-LAT $\gamma$-ray data for a large sample of bright  core radio MAGN.
These results support the hypothesis that $\gamma$-ray  photons are produced by synchrotron-self Compton scatterings 
by the same electrons responsible for the synchrotron emission at  GHz frequencies. 
Ref.~\cite{RG} quantifies several sources of uncertainty  and estimates them to affect the diffuse emission from unresolved MAGN by about a factor of 10. 
 The expected  flux is therefore provided as a band  (roughly decreasing with $E^{-2.35}$) affected by the extragalactic 
 background light (EBL) absorption above 50 GeV. 
At 1 GeV, the MAGN flux (multiplied by $E^2$) ranges from 6$\cdot 10^{-7}$  GeV cm$^{-2}$ s$^{-1}$ sr$^{-1}$ to 
5$\cdot 10^{-6}$  GeV cm$^{-2}$ s$^{-1}$ sr$^{-1}$, 
while at 100 GeV the EBL absorption starts to be effective and the flux spans from 9$\cdot 10^{-9}$  GeV cm$^{-2}$ s$^{-1}$ sr$^{-1}$ 
to 9$\cdot 10^{-8}$  GeV cm$^{-2}$ s$^{-1}$ sr$^{-1}$.
\\
The result in Ref.~\cite{RG} states that MAGN account for at least 10\% of the IGRB measured by {\it Fermi}-LAT \cite{IDGRB} and can in principle explain most of it.  
Added to other possible diffuse emission components, this might almost close the room to an exotic diffuse contribution such as from DM annihilation in space.
 This point is the focus of our paper: evaluating how much the MAGN $\gamma$-ray
diffuse emission constrains the presence of WIMP DM  annihilating  in the galactic halo. In the  remainder of this Section, 
we briefly discuss the astrophysical sources leading to a non-negligible diffuse $\gamma$-ray emission at  $|b| > 10^{\circ}$ 
(see Ref.~\cite{2012PhRvD..85b3004C} for further details). 
\\
{\it Blazars.}
The census of the sources detected by the {\it Fermi}-LAT \cite{1FGLC, 2FGLC} indicates that the largest population is made by 
blazars, a class of AGN with their jet oriented along the l.o.s., thus leading to a significant observed $\gamma$-ray flux.
The analysis of the spectral and statistical properties of the resolved {\it Fermi}-LAT blazars \cite{2010ApJBlazarFermi} 
shows that the unresolved counterpart  is a sizable component, able to 
 explain up to  $20\% - 30\%$ of the IGRB.  The contribution from the two distinct populations of Flat Spectrum Radio Quasars 
(FSRQs) and BL Lacertae (BL Lacs) objects has been found to follow a power-law spectrum with respectively a somewhat softer 
 ($\alpha_{FSRQs} = 2.45 \pm 0.03 $) and harder ($\alpha_{BL Lacs} = 2.23 \pm 0.03 $) spectral index.
Recently, Ref.~\cite{2012ApJ...751..108A} has evaluated the FSRQ contribution to the diffuse $\gamma$-ray flux using a new 
determination of the spectral energy distribution and luminosity function. The resulting flux is predicted with an uncertainty 
of less than a factor of two. It represents no more than 10\% of the IGRB in the 0.1-100 GeV, and becomes negligible for E $\gsim$ 20 GeV.
Also, \cite{2012JCAP...11..026H} and \cite{2012PhRvD..86f3004C} find that the maximal contribution from blazars are about 10\% and 20\%,
respectively. 
\\
{\it Milli-Second Pulsars.} 
The second-most abundant population of the {\it Fermi}-LAT sky is represented by galactic pulsars and 
Milli-Second Pulsars (MSPs). It has been argued that the latter population 
may be brighter in $\gamma$-rays than ordinary pulsars as a consequence of
 their rapid spin frequency \cite{2005ApJ...622..531H}. Moreover, MSPs are expected to be more numerous 
 at high latitudes, since they have typical ages that can largely exceed the oscillation time across the galactic disk \cite{FG2010}. 
 The prediction of the diffuse $\gamma$-ray flux due to high latitudes ($|b|>10^{\circ}$) unresolved MSPs  is estimated to be about  
$10 \%$ at $\simeq 3 \,  {\rm GeV}$, where it reaches its maximum value due to the adopted broken power-law spectrum 
 \citep{2012PhRvD..85b3004C}.  
 The considered model results to be compatible with the analysis of the $\gamma$-ray anisotropies performed in 
   \cite{Ackermann:2012uf}. Indeed we derived the integrated flux from MSPs for $|b| >30^{\circ}$,  which results to be 1.44\% and 1.11\% 
   of the IGRB in the 1-2 GeV and 2-5 GeV, respectively. These numbers are compatible with the upper 
   limits of about 1.5\% found by \cite{Ackermann:2012uf} in the anisotropy  analysis.
 In the Second {\it Fermi} catalog \cite{2FGLC} only about 20 MSPs at  $|b| > 10^{\circ}$  are listed, but the number of galactic 
 (non-globular cluster) MSPs has recently risen to 40 \cite{2013arXiv1305.4385T}. The statistical properties of such a population 
 could improve the prediction for the unresolved component. Such an analysis is beyond the scope of the present paper,  but is anyway 
 not expected  to modify any of the results and conclusions presented here. 
In the following, we refer to the unresolved MSPs diffuse $\gamma$-ray flux and its estimated 
uncertainty as reported in Ref.~\cite{2012PhRvD..85b3004C}.
\\
{\it Star-forming galaxies.} 
Though their number density is larger than that of the blazars, star-forming (SF) galaxies are more difficult to 
detect in $\gamma$-rays because of their less intense luminosities.
Based on three years of {\it Fermi}-LAT data, Ref.~\cite{2012ApJ...755..164A} analyzes a sample of 64 SF galaxies beyond the Local Group, 
detecting only four objects in $\gamma$-rays: two are typical star-burst galaxies (M82 and NGC 253), 
while the other two host Seyfert 2 nuclei (NGC 1068, NGC 4945). 
None of the remaining 60 source candidates has shown an excess above the background.
The contribution of SF galaxies to the IGRB is evaluated in  Ref.~\cite{2012ApJ...755..164A} by exploiting the 
correlation between infrared and $\gamma$-ray luminosities   \cite{2011ApJ...736...40S}. 
 In order to compute the diffuse emission, the differential photon flux of an individual galaxy has to be modeled. 
 Given the lack of statistics, the $\gamma$-ray average 
 spectral properties of SF galaxies are difficult to firmly establish, although differences in the spectra of quiescent and star-burst galaxies are expected to 
 be present. To take into account this effect, Ref.~\cite{2012ApJ...755..164A} adopts two possible $\gamma$-ray 
 spectra supposed to bracket the possible contribution 
 from multiple types of SF galaxies. The first one refers to Milky Way-like SF galaxies (model MW), the second one assumes a power-law spectrum in
  agreement with the properties of the {\it Fermi}-LAT detected star-burst galaxies (model PL). 
The predictions differ in particular above 5 GeV, where the MW model gets significantly softer. At 100 GeV the PL model overpredicts the MW one by one 
order of magnitude.  The estimates span 4\%-23\% of the IGRB intensity above 100 MeV.
Higher predictions, in particular in the low energy tail (not relevant in deriving the constraints 
on the DM annihilation cross section),  have been obtained in \cite{2010ApJ...722L.199F,2013ApJ...773..104C}. 
\\
As a summary of this Section, we show in Fig.\,\ref{fig:fluxes} the contributions to the IGRB discussed above, 
with uncertainty bands as reported in the relevant references (roughly corresponding to $1\sigma$ deviations).
The {\it Fermi}-LAT IGRB data \cite{IDGRB} are shown along with the power-law fit (solid black line).
The solid dark-green curve refers to the expected diffuse $\gamma$-ray flux from unresolved MAGN  \cite{RG}, the light-green band bracketing the corresponding 
uncertainty. The other displayed components are: SF galaxies (the MW model is depicted by the dashed dark- (light-) blue line (area)), 
MSPs (dot-dashed red line and orange band), 
FSRQs (purple dotted line and pink band), and BL Lacs (dotted dark-grey line and light-grey band).
\\
The BL Lacs flux follows a power-law, while the specific shape of the FSRQ contribution 
is the result of a more recent, sophisticated analysis of their spectral energy distribution and luminosity function \cite{2012ApJ...751..108A}. 
The crossing point for the two curves is around 1-2 GeV: above this energy the BL Lacs flux 
dominates over the one from FSRQs. 
 The $\gamma$-rays from 
unresolved MSPs show a peculiar spectrum peaked at about 1 GeV (resulting from the primary 
electron cooling due to curvature radiation) and dominate over the blazar spectra from 300 MeV up to 3-4 GeV. 
A similar low-energy \textsl{bumped} flux comes from the SF galaxies contribution. The MAGN is clearly the dominant component over the whole 
energy range (notice that for $E_\gamma\gtrsim30\,$GeV the BL Lacs 
flux is actually expected to drop in the same way as the MAGN flux, due to EBL absorption, 
but this has not been taken into account in the flux estimate referred to here).
In the following, we define  $\Sigma_{\rm BMS}$ as the sum of the contributions 
from Blazars (both BL Lacs and FSRQs), MSPs and SF galaxies.  

\begin{figure}[t!]
\includegraphics[scale=0.65]{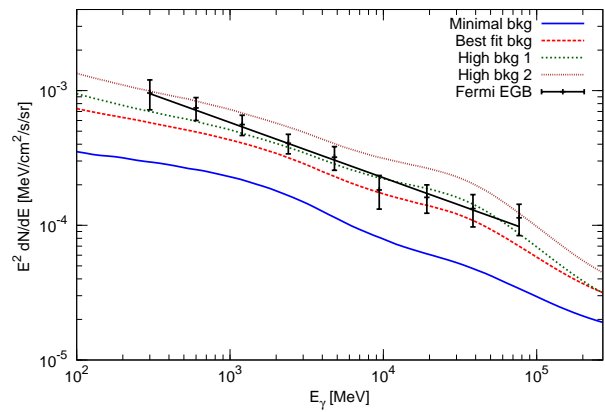}  
\caption{Different representative choices for the astrophysical backgrounds
shown in Fig. \ref{fig:fluxes}. The solid blue (red dashed) line corresponds to the minimal (best fit) 
prediction for all the contributions. The green short-dashed (brown dotted) curve has been obtained from the 
sum of the best fit models for $\Sigma_{\rm BMS}$ and the 60\% (85\%) of the maximally expected flux from MAGN.
Data points are again for the IGRB \cite{IDGRB}. } 
\label{fig:backgrounds}  
\end{figure}

In Fig. \ref{fig:backgrounds} we display a few representative choices for the 
astrophysical backgrounds, as a function of the photon energy. 
The solid blue (red dashed) line corresponds to the minimal (best fit) 
prediction for all the contributions. The green short-dashed (brown dotted) curve has been obtained from the 
sum of the best fit models for $\Sigma_{\rm BMS}$ and 60\% (85\%) of the maximally expected flux from MAGN.
We also display the data for the IGRB \cite{IDGRB} with 1$\sigma$ error bands.

These figures illustrate that by varying the contribution of each component within its uncertainty band, the IGRB data can be saturated 
by multiple combinations of the above fluxes. Alternatively, only little room is left to other unknown $\gamma$-ray sources as, \textsl{e.\,g.}, DM.

\section{Diffuse $\gamma$-ray emission from Dark Matter in the galactic halo}
\label{sect:DM}
The self-annihilation of DM particle pairs in the haloes of galaxies may give birth, among other species, to
$\gamma$-rays. The intensity of these energetic photons depends on the 
elementary process at stake and on the spatial distribution of DM. 
The flux  $\Phi_\gamma(E_\gamma, \psi)$ of $\gamma$-rays produced by
WIMP pair annihilation in the angular direction $\psi$ is given by \cite{1990NuPhB.346..129B,1998APh.....9..137B,Bottino:2004qi} :
\begin{equation}
\Phi_\gamma(E_\gamma, \psi) = \frac{1}{4\pi} \frac{\langle\sigma v\rangle}{m_\chi^2} \frac{dN_\gamma}{d E_{\gamma}} \frac{1}{2}I(\psi), 
\label{eq:flux_gamma}
\end{equation}
where $m_\chi$ is the WIMP mass. The factor  \sigmav defines the annihilation cross section times the relative
velocity, averaged over the galactic velocity distribution function, while
$dN_\gamma/d E_{\gamma}$ is the $\gamma$-ray production source spectrum 
 per DM annihilation event.
The last term  $I(\psi)$ is the integral performed along the l.o.s.\,of the squared DM density distribution:
 \begin{equation}
 \label{Ipsi} 
 I(\psi) = \int_{l.o.s.}\rho^{2}(r(\lambda, \psi))d\lambda\,.
\end{equation} 
Here  $\psi$ is the angle between the l.o.s.\,and the direction towards the 
galactic center, defined as a function of the galactic latitude $b$ and longitude $l$ 
($\cos\psi = \cos b\cos l$). When comparing with experimental data, Eq.\,(\ref{Ipsi}) must be 
averaged over the telescope viewing solid angle, 
$\Delta\Omega$: 
 \begin{equation}
 \label{I_omega}
I_{\Delta\Omega} = \frac{1}{\Delta\Omega}\int_{\Delta\Omega}I(\psi(b, l))d\Omega\,. 
\end{equation}
The photon spectrum $dN_\gamma/d E_{\gamma}$ depends on the elementary 
processes ruling the annihilation. For the continuum
$\gamma$-ray flux, we consider here $i)$ prompt emission, 
where the photons are found in final-state showers or hadronic decays of the annihilation products,
and $ii)$ ICS by energetic electrons and positrons --  produced in the same way -- off the interstellar radiation field (ISRF). 
We have calculated  the  spectrum for the prompt emission of both photons and $e^\pm$ with the Pythia Montecarlo code (version 8.162) \cite{pythia}.  
\\
The ICS has been computed from those $e^\pm$  interacting with ambient photons, following the prescription in 
\cite{2009NuPhB.821..399C}.   
Once produced, energetic electrons may diffuse due to stochastic scatterings on  galactic magnetic field inhomogeneities, and loose energy 
through interactions with the ISRF. The ISRF is composed by
the Cosmic Microwave Background (CMB) photons, the infrared radiation (IR) produced by the absorption and the subsequent re-emission of starlight by the 
galactic dust, and by the starlight (SL) originating from stars of the Galactic disk. 
Given the energies at stake, 
galactic diffusion may be safely neglected and only energy losses have been included, as usually assumed in the literature \cite{Regis:2008ij}. 
Electromagnetic energy losses are treated here in the fully relativistic Klein-Nishina regime. 
The photon density distribution has been assumed as an average field with different normalizations for different sky regions
(the SL and IR are concentrated along the galactic disk), following Ref.~\cite{2009NuPhB.821..399C}. 
Specifically, between 10$^\circ$ and 20$^\circ$ all the three fields (CMB, IR, SL) are present, while for latitudes 
$|b|>20^\circ$ only CMB remains. 
Energy losses due to synchrotron radiation have been taken into account in the region of latitude $10^\circ<|b|<20^\circ$, with a value of $B=2\,  \mu$G for the galactic magnetic field,
and neglected above $|b|>20^\circ$ given the very low intensity of $B$ at high latitudes \cite{2008A&A...477..573S,1994A&A...288..759H,2003A&A...410....1P,2011ApJ...738..192P}.
Switching off the synchrotron radiation would increase the ICS $\gamma$-ray flux from DM
 by less than 30\%, resulting in a decrease of the upper limits on \sigmav 
(shown in the following Sect. \ref{sect:constraints})  by 10\%.

Numerical cold DM simulations predict cuspy density distribution in the inner parts of galaxies, like the 
Navarro-Frenk-White (NFW) \cite{Navarro:1995iw} or Einasto \cite{1965TrAlm...5...87E} profile. Following a conservative approach, however, 
we evaluate the  geometrical factor $I(\psi)$  in Eq. \ref{Ipsi}
 by adopting the cored Burkert model \cite{1995ApJ...447L..25B} for the radial DM density distribution, 
which is  used as an alternative parameterization to fit observed rotation curves \cite{2013JCAP...07..016N}:
\begin{equation} 
\label{rhoburk}
   \rho(r)=\frac{\rho_{s}}{\left(1+ \left( \frac{r}{r_s} \right)^2\right)\left(1+ \frac{r}{r_s}\right)}\;.
\end{equation}
We note that numerical simulations typically support more cuspy profiles and our choice is
thus exclusively motivated by that fact that it results in rather conservative limits on a DM contribution to the IGRB.
 The values of $r_{s}$ and $\rho_{s}$ have been derived fixing the local DM density
 $\rho(r\!=\!r_{\odot})=0.4$ GeV cm$^{-3}$ \citep{2010JCAP...08..004C,2010A&A...523A..83S}
 ($r_\odot$=8.33 kpc) and fitting the halo DM mass $M(r<r_0)$
contained within a certain radius $r_0$. 
Using available $M(r<r_0)$ data for different values of $r_0$ \cite{2012MNRAS.425.2840D,2010ApJ...720L.108G,2013A&A...549A.137I,2012ApJ...761...98K,1996ApJ...457..228K,2003A&A...397..899S,2009MNRAS.398.1757W,1999MNRAS.310..645W,2008ApJ...684.1143X},
 roughly in the range $20\,{\rm kpc}\lesssim r_0\lesssim 300\,{\rm kpc}$, we find a best fit value of $r_s = 15$ kpc.
Numerically, we derive for $I_{\Delta\Omega}$ in Eq.\,(\ref{I_omega}):  2.4, 1.9, 1.7, 2.0 (in units of GeV$^2$ cm$^{-6}$ kpc)  
for the sky regions $ 10^{\circ} < |b| < 20^{\circ}$, $ 20^{\circ} < |b| < 90^{\circ}$, $|b|>60^{\circ}$, and $|b|>10^{\circ}$ respectively.  
While in principle limits might somewhat improve when taking into account the different values of $I_{\Delta\Omega}$ for different regions of the sky, 
we will in the following only consider the whole region $|b|>10^{\circ}$ considered by the {\it Fermi}-LAT analysis \cite{IDGRB} because not all of the 
astrophysical contributions quoted above can be easily rescaled to smaller subregions.
We also note that when integrating Eq.\,(\ref{Ipsi}) around the galactic center, 
 different DM  distributions $\rho(r)$ may lead to very different results. However, since we study here only
high latitude regions, the various models for $\rho(r)$ give comparable values for $I(\psi)$, 
and the uncertainty due to DM density modeling  is rather small. For example,  we find  
$I_{\Delta\Omega} (|b|\!>\!10^{\circ})\cdot{\rm GeV}^{-2}{\rm cm}^{6}{\rm kpc}^{-1}=$ 2.4 (2.5, 2.8) for the case of an isothermal sphere
 (a NFW  \cite{Navarro:1995iw}, an Einasto  \cite{1965TrAlm...5...87E}) profile.  
While the existence of a dark disk \cite{Read:2008fh} may further enhance those values slightly, we will not consider this possibility in the following.
We also do not include any  contribution from sub-haloes, which are predicted in numerical simulations of cosmological
structure formation \cite{2011ASL.....4..297D,2008MNRAS.391.1685S}, and assume a smooth distribution of DM in the galactic halo
for the sake of deriving conservative limits. Including the effect of sub-structures leads to an increase of the geometrical factor in Eq.\,(\ref{I_omega}), 
see e.g.~Ref.~\cite{Kamionkowski:2010mi}, which would accordingly lower the upper limits we are going to 
 present on \sigmav\!.

\section{Constraints on Dark Matter annihilation into $\gamma$-rays}
\label{sect:constraints}

The constraints on DM are set, as usual,  in terms of the velocity averaged 
annihilation cross section into $\gamma$-rays \sigmav as a function of the WIMP mass 
$m_\chi$. All the other quantities in Eq.\,(\ref{eq:flux_gamma}) are kept fixed.  
The galactic DM contribution is calculated following the prescriptions outlined in Sect.\,\ref{sect:DM}.
We do not include the diffuse $\gamma$-ray flux from  DM 
 in extragalactic structures \cite{2001PhRvL..87y1301B,2002PhRvD..66l3502U}, given the   large model-dependence on the assumed DM (sub-)halo profiles and their evolution. 
 If the $\gamma$-rays from DM in  extragalactic structures or sub-haloes in the Milky Way halo  were added to our fluxes, 
 the ensuing upper limits  on \sigmav would be further decreased. 
In these regards, the results we are going to present are meant 
to be conservative, $i.e.$\,leading to the highest possible value for the excluded $\langle\sigma v\rangle$.

As discussed in Sect.\,\ref{sect:DM}, the geometrical factor at high latitudes is very slightly affected by modeling uncertainties.
The choice we made on the Burkert profile and a smooth halo  is meant to be conservative and different assumptions on the 
DM profile would  lower the upper bounds we find by up to about 30\% 
(in fact, considerably more for optimistic assumptions about DM substructure \cite{Diemand:2006ik,Kuhlen:2008aw,Pieri:2009je}).
A further uncertainty factor is the value of the
 local DM density, which can span the (generous) range 0.2 - 0.7 GeV cm$^{-3} $ \cite{2010A&A...523A..83S}. 
 Since $\rho_\odot$ enters squared in the flux, it has a non-negligible impact on the \sigmav bounds.
  If not explicitly  stated differently, we will take $\rho_\odot$=0.4 GeV cm$^{-3}$ in the following.  
  To keep the analysis as model independent as possible, we 
will show our results assuming that DM annihilation proceeds via a single channel (branching ratio 
equal to one), studying separately the effect of each final state leading to $\gamma$-ray production.

Finally, we derive conservative limits on \sigmav by requiring that the total gamma-ray flux does not exceed any of the 
{\it Fermi}-LAT data points \cite{IDGRB} by more than  $n\sigma$, which is a commonly adopted practice for this kind of constraints (see e.g. Ref.~\cite{Abdo:2010dk}). 
Our limits are thus set by  the maximal \sigmav value for which 
$\Phi_i^{\rm tot} \leq \Phi_i^{\rm exp} + n \times \Delta(\Phi_i^{\rm exp})$ in all energy bins $i$,
where $\Phi_i^{\rm tot}$ is the sum of the DM and astrophysical background fluxes in a given bin $i$, 
and $\Phi_i^{\rm exp}$ is the experimental data with error $\Delta(\Phi_i^{\rm exp})$. 
Given the one-sided nature of the limit, a 2$\sigma$ (3$\sigma$) upper limit
from the intensity data points thus nominally corresponds to a confidence level of 97.7\% (99.9\%).
We typically quote $2\sigma$ limits -- except for the case of a large
 MAGN contribution, c.f.~the brown dotted line in Fig.~\ref{fig:backgrounds}, where we will instead consider the more conservative limits resulting from the requirement that no 
 data point is exceeded by more than $3\sigma$.\footnote{
 Note that in such a situation, where the signal must obviously be very small, the upper limit on 
 \sigmav that is inferred from an analysis like ours may be anomalously low (e.g.~if the data `fluctuate very low relative to the expectation of 
 the background-only hypothesis' \cite{Beringer:1900zz}). For a general introduction to this problem see Section 36.3.2.2 of the 
 Particle Data Group review \cite{Beringer:1900zz}, or Ref.~\cite{Junk:1999kv,Read:2000ru} for more details.
 }
 As can be anticipated from Fig.~\ref{fig:fluxes},  those limits are always set by the 
 three data bins at 6.4--12.8, 12.8--25.6 and 25.6--51.2 GeV.

\begin{figure}[t!]
\includegraphics[scale=0.65]{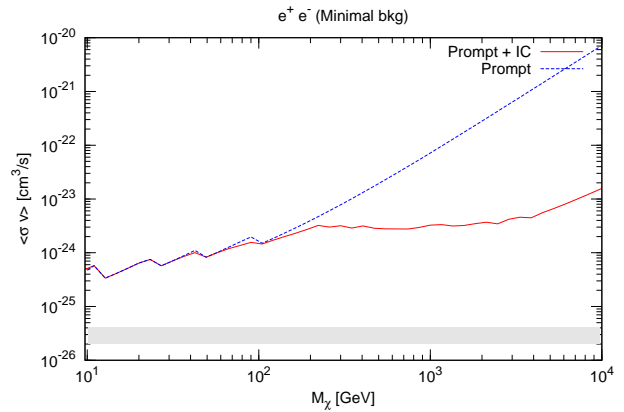}  
\caption{Conservative 2$\sigma$ upper limits on the velocity averaged DM annihilation cross section  \sigmav
 into $e^\pm$ final states, as a function of the WIMP mass  $m_\chi$. 
 The blue dashed curve corresponds to the  $\gamma$-ray spectrum from the prompt component  only, the solid red line refers to the inclusion of ICS.  
 The grey area indicates the typically adopted 
 value for the cross section of thermally produced WIMPs, $\langle\sigma v\rangle\sim3 \cdot 10^{-26}$ cm$^{3}$ s$^{-1}$.} 
\label{fig:boounds_theorunc_ee}  
\end{figure}

\begin{figure*} [t!]
\includegraphics[width=\columnwidth]{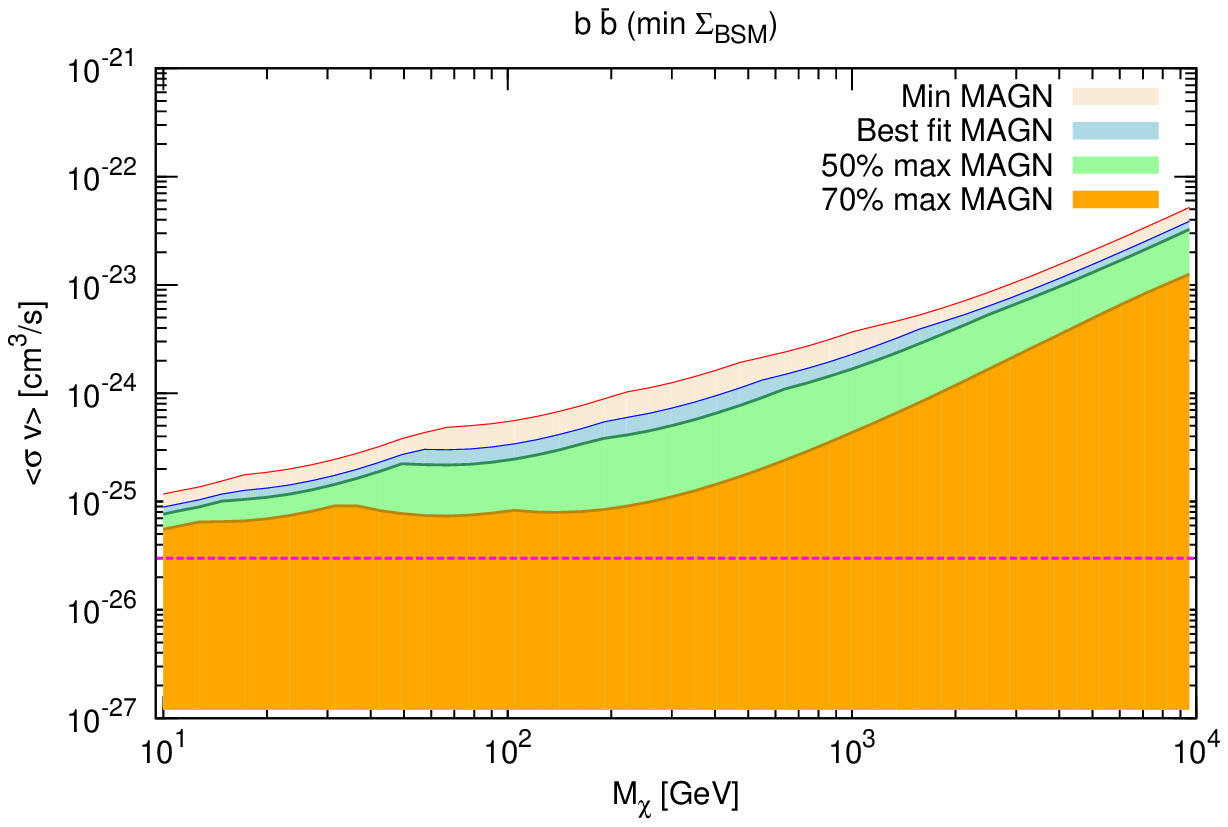} 
\includegraphics[width=\columnwidth]{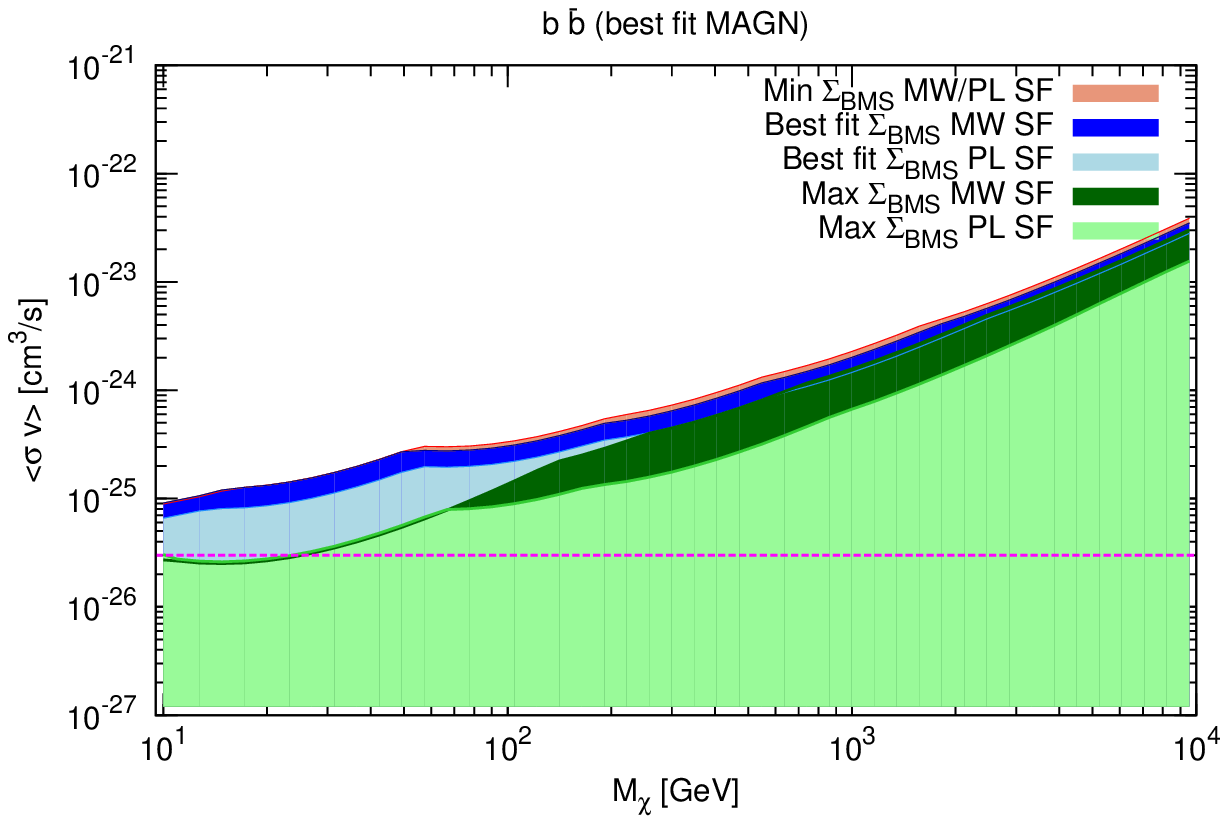} 
\caption{2$\sigma$  upper limits on the velocity averaged DM annihilation cross section  \sigmav
 as a function of the WIMP mass, for different levels of $\Sigma_{\rm BMS}$
and MAGN. The parameters in the colored areas are allowed by $\gamma$-ray data and 
astrophysical backgrounds   by the  {\it Fermi}-LAT IGRB data \cite{IDGRB}.  
 In the left panel, we fix $\Sigma_{\rm BMS}$ at the minimal
 level compatible with the expectations and vary the contribution from MAGN from the minimal 
to 70\% of the maximally expected level.
 The right panel shows the impact of various choices for $\Sigma_{\rm BMS}$ on the limits, fixing the MAGN contribution 
to the best-fit value derived in Ref.~\cite{RG}.
 The constant line at $3 \cdot 10^{-26}$ cm$^{3}$ s$^{-1}$ represents the 'thermal' cross section of DM in the simplest WIMP  models.
\label{fig:limits1} 
} 
\label{fig:bounds_sf} 
\end{figure*}

Fig.\,\ref{fig:boounds_theorunc_ee} represents the exclusion plot for DM annihilation into $e^{\pm}$ and it is emblematic of the effect of the
ICS inclusion in the DM spectrum. The blue dashed curve corresponds to upper limits on \sigmav when the DM induced spectrum 
originates only from the prompt $\gamma$-ray component, while the solid red line refers to the inclusion of ICS. 
The astrophysical background is set 
to the minimal choice illustrated in Fig.\ref{fig:backgrounds}.
The effect of ICS inclusion starts to be important for high energies, \textsl{i.\,e.\,}at high DM masses, 
and the deviation from the \textsl{prompt only} limits reaches 
about  two orders of magnitude at $\sim$ 10 TeV.
We notice that the uncertainties from ICS are small at the energies at which it is effective. 
 For comparison, we also indicate in the same plot the 'thermal' cross section of  
 $\langle\sigma v\rangle\sim3 \cdot 10^{-26}$ cm$^{3}$ s$^{-1}$ that results in the correct DM relic density in the 
 simplest WIMP scenarios (i.e.~velocity-independent annihilation and no co-annihilations with other particles close in mass to the WIMP).

\begin{figure*} [t!]
\includegraphics[width=\columnwidth]{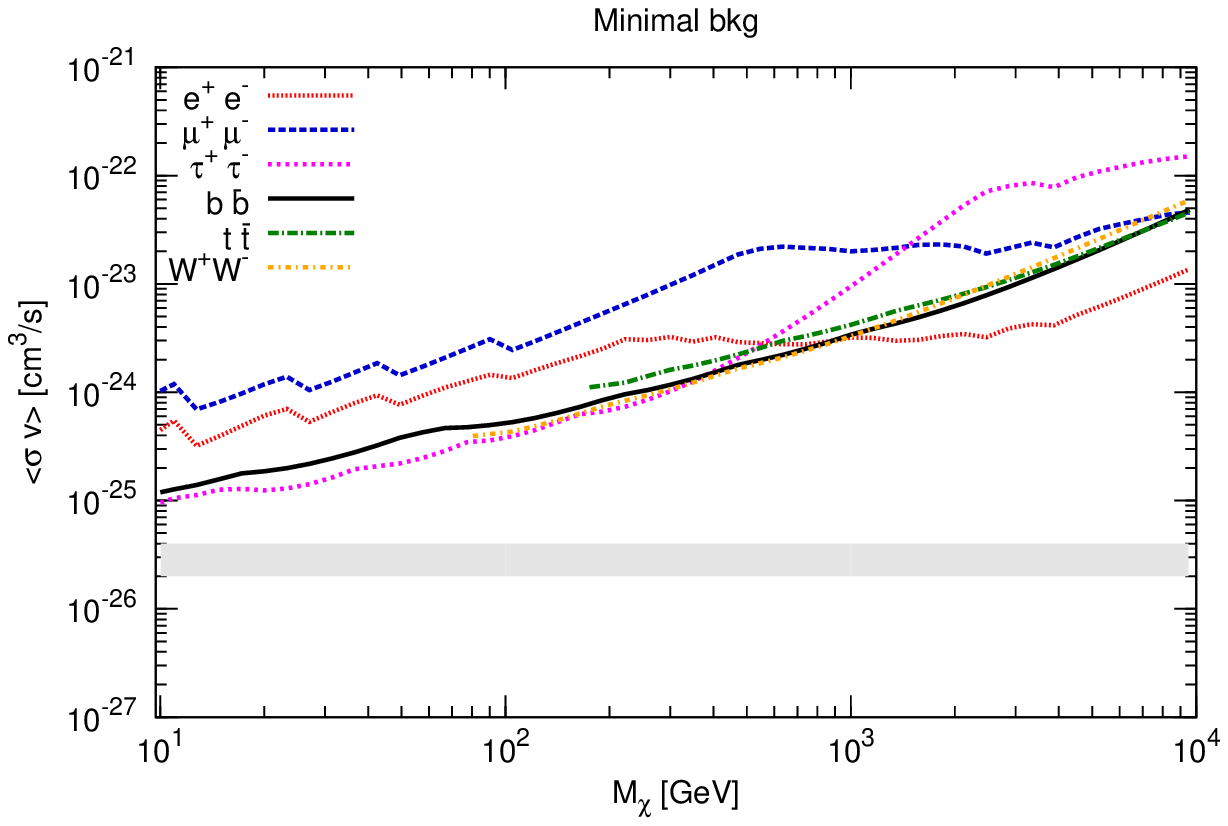} 
\includegraphics[width=\columnwidth]{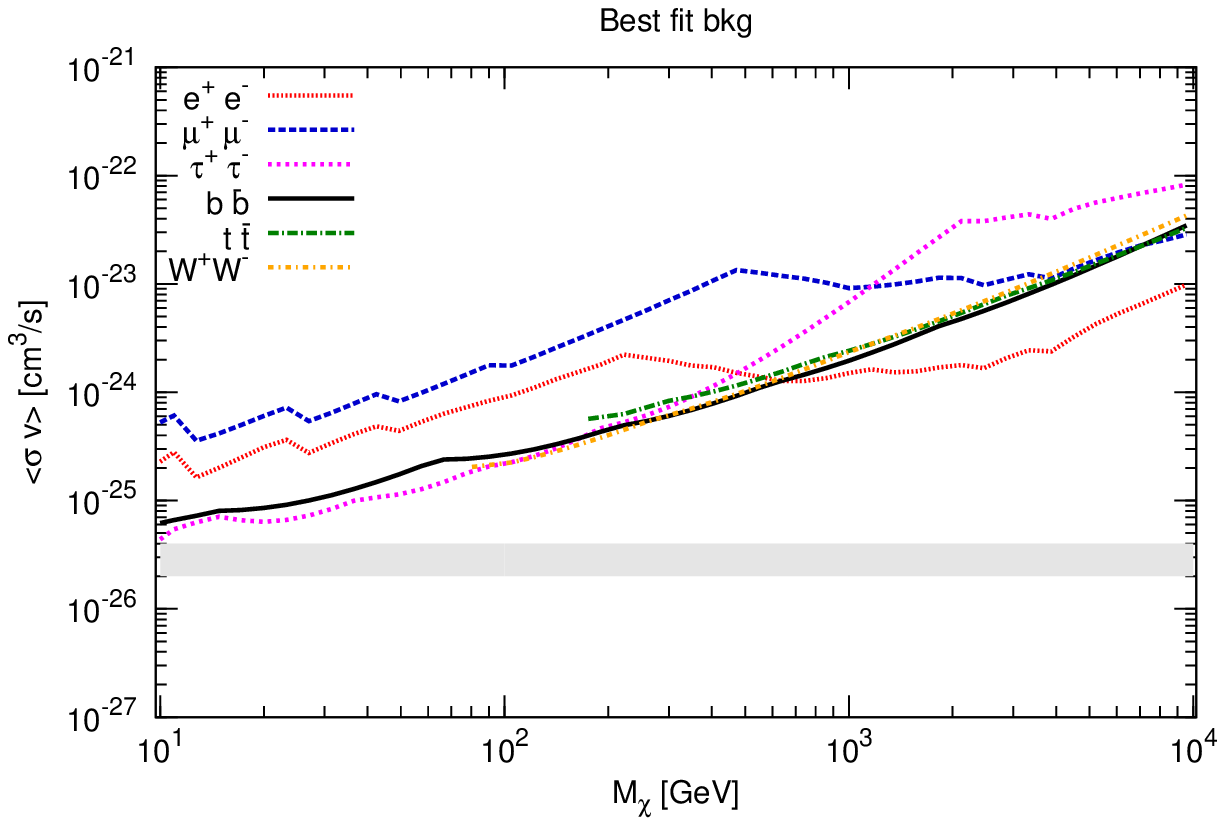} 
\\
\includegraphics[width=\columnwidth]{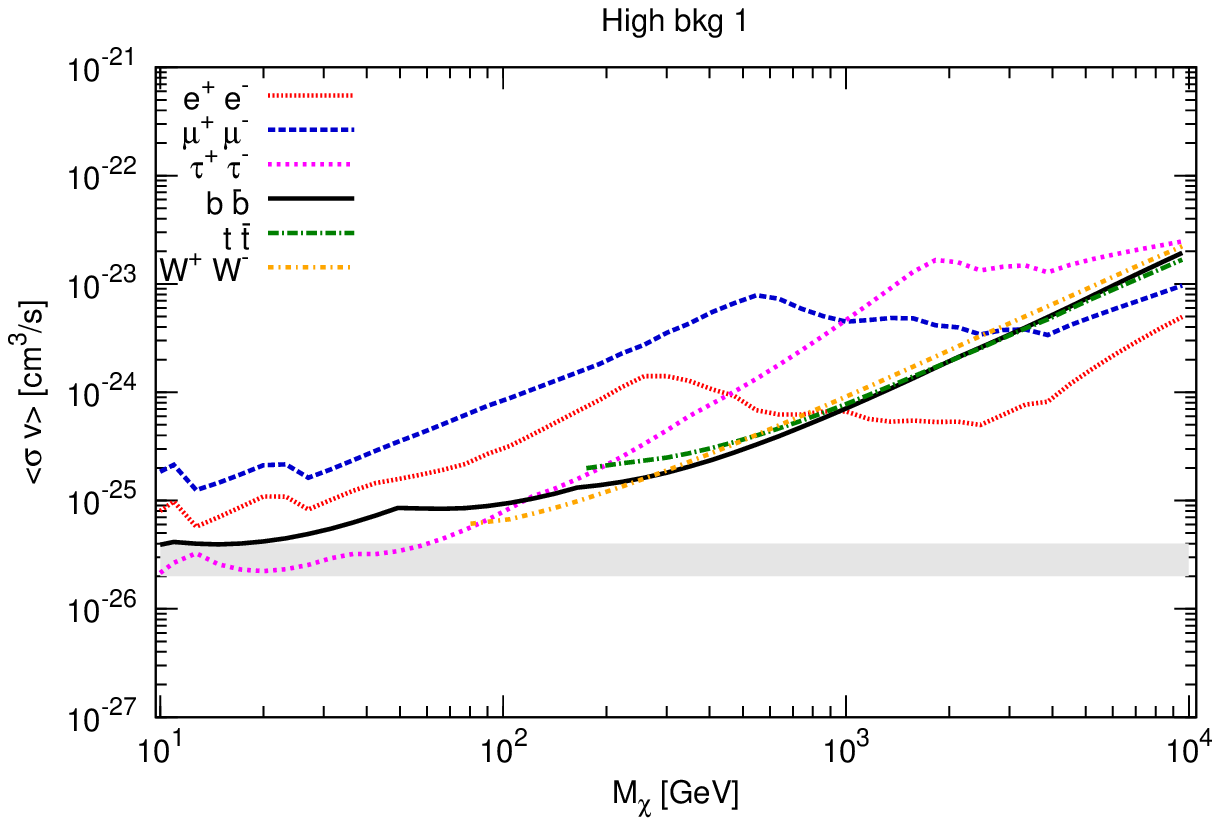} 
\includegraphics[width=\columnwidth]{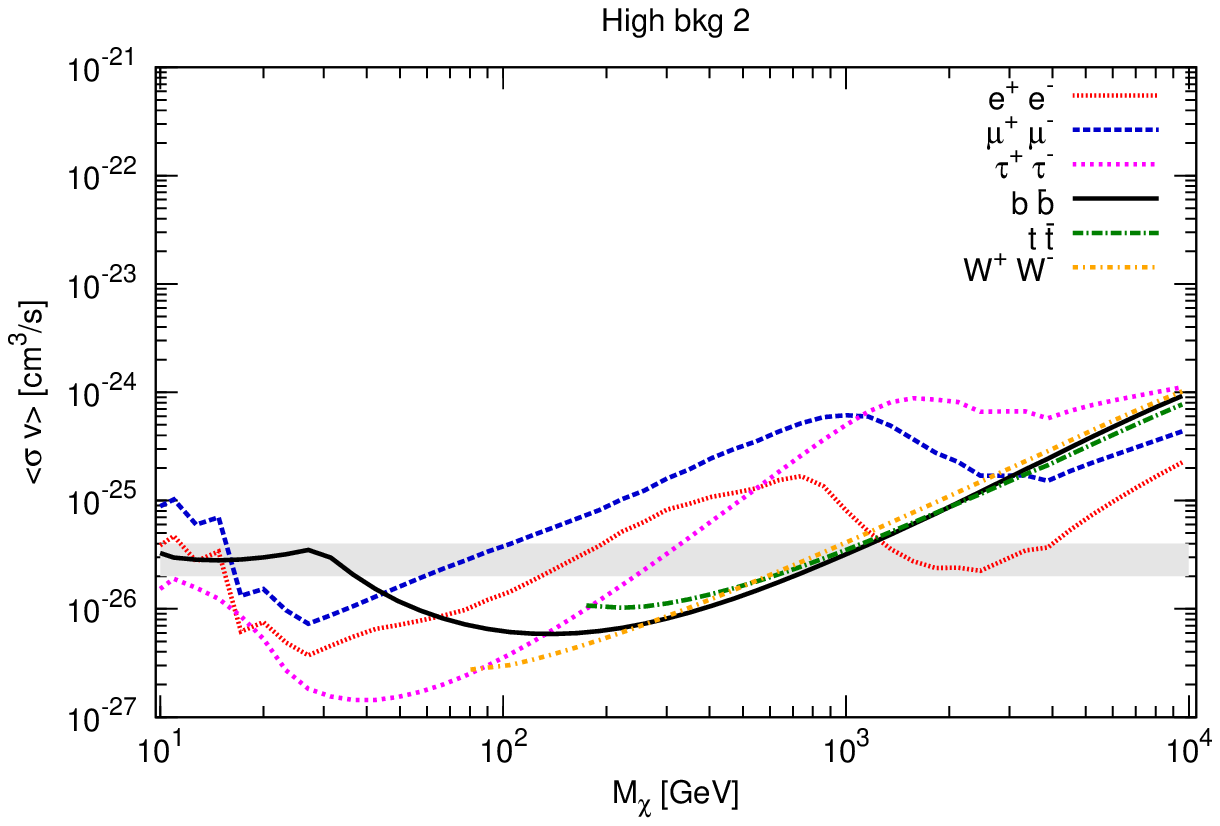} 
\caption{ Upper limits on the velocity averaged DM annihilation cross section  \sigmav into $\gamma$-rays from the {\it Fermi}-LAT 
IGRB data \cite{IDGRB}, for various possible annihilation channels. The top left (right) panel shows our most conservative (more realistic) limits, 
obtained by setting all astrophysical contributions to the minimally allowed ('best-fit') level. 
The bottom panels, where we also fix $\Sigma_{\rm BMS}$ at the 'best-fit' value, represent a more optimistic choice for the MAGN contribution 
concerning potential DM limits: 
in the left (right) panel MAGN are assumed to actually contribute 60\% (85\%) of the maximally expected flux. Note that for the limits in the
bottom right panel we require the total $\gamma$-ray flux not to overshoot any data point by $3\sigma$ (rather than $2\sigma$ like in all 
other cases presented in this work).
The grey area indicates the typical benchmark value for the cross section of thermally produced 
WIMPs, $\langle\sigma v\rangle\sim3 \cdot 10^{-26}$ cm$^{3}$ s$^{-1}$.
\label{fig:upperlimits}
}
\end{figure*}

In Fig.\,\ref{fig:bounds_sf} we study the effect of the various astrophysical backgrounds described in Sect.\,\ref{sect:IGRB}
on the room left for a possible DM contribution to the diffuse $\gamma$-ray flux,
studying in particular the effect of MAGN and SF galaxies. The annihilation channel is fixed to $b\bar{b}$. 
In the left panel, we study the impact of the previously neglected MAGN component by fixing 
$\Sigma_{\rm BMS}$ to the minimally allowed value: by increasing the MAGN contribution from its minimal  
to 70\% of its maximally expected value \cite{RG}, limits on a possible DM contribution could be improved 
by at least one order of magnitude compared to the most conservative assumptions concerning the astrophysical 
contributions (note that setting the $\gamma$-ray flux from MAGN to the maximally allowed value would actually 
overshoot the {\it Fermi} data by more than $2\sigma$). 

The right panel of Fig.\,\ref{fig:bounds_sf}, on the other hand, demonstrates that the uncertainty in the flux 
from other astrophysical backgrounds generally has a less significant impact on the final limits. Here, we fix the 
MAGN contribution to its best-fit value \cite{RG} and vary $\Sigma_{\rm BMS}$ from its minimally to maximally
predicted contribution as described in Section \ref{sect:IGRB}. We also show that the choice of the underlying 
SF galaxy model introduces a certain systematic uncertainty: depending on the WIMP mass, the resulting limits are
up to a factor of $\sim$3 stronger for the PL than for the MW model (unless  $\Sigma_{\rm BMS}$ only contributes 
at the minimal level, in which case there is no difference because the MAGN contribution always dominates).

In Fig.~\ref{fig:upperlimits}, we show the effect of the MAGN contribution for various DM annihilation channels.  
In the top left panel, we present our most conservative limits, 
derived by fixing the background to the minimal scenario (see Fig. \ref{fig:backgrounds}). 
For $b\bar{b}$, $\mu^+\mu^-$ and $\tau^+\tau^-$ 
final states, and DM masses below $\sim$100\,GeV,  they are comparable with the limits previously derived in 
Ref.~\cite{2012PhRvD..85b3004C}, where a similar modeling of the astrophysical background was considered; 
for $m_\chi\gtrsim100$\,GeV the inclusion of ICS becomes important and consequently significantly improves 
our new limits with respect to those for leptonic channels. The right panel shows a more realistic scenario, 
where we take all astrophysical backgrounds to contribute at their expected 'best-fit' values
(dashed red curve in Fig. \ref{fig:backgrounds}).
Compared to the conservative case, this shifts down the upper bounds by a factor of roughly 2.

In the lower panels of Fig.~\ref{fig:upperlimits}, we take a more optimistic approach in order to illustrate 
the full potential of using the IGRB to constrain galactic DM annihilation. Assuming MAGN to contribute at 60\% 
of the maximal flux (high background 1, corresponding to the short-dashed green curve in Fig. \ref{fig:backgrounds}), for example, 
would result in limits that clearly start to approach the 'thermal' 
cross section of  $\langle\sigma v\rangle\sim3 \cdot 10^{-26}$ cm$^{3}$ s$^{-1}$. Moving to the possibility of 
even higher levels for the MAGN contribution, we note that one quickly reaches a situation where already a small 
increase in the MAGN flux has an enormous impact on the allowed DM contribution -- simply because the astrophysical 
backgrounds are already extremely close to the IGRB data points (with 2$\sigma$ errors). In this limit, the validity 
of our method to obtain upper bounds  clearly breaks down and a more sophisticated statistical treatment would be needed. 
In order to illustrate that one may still potentially be able to constrain the 'thermal' annihilation rate up to DM
masses of several TeV,  we  show in the bottom right panel of Fig.~\ref{fig:upperlimits} the limits that result for an assumed 
MAGN contribution at 85\% of the maximally predicted flux
(high background 2, corresponding to the dotted brown curve in Fig. \ref{fig:backgrounds}).
Note that those limits were derived by requiring that no data point 
is exceeded by more than 3$\sigma$, rather than 2$\sigma$, in order to (partially) meet the above concerns.

Let us stress in passing that the inclusion of ICS $\gamma$-rays is particularly relevant for large DM masses ($m_\chi\gtrsim1$\,TeV). This is the mass range 
which -- for leptonic final states \cite{Bergstrom:2009fa,Yuan:2013eja,Cholis:2013psa,Jin:2013nta} -- could in 
principle contribute to the observed rise in the cosmic ray positron fraction \cite{Adriani:2008zr,FermiLAT:2011ab,Aguilar:2013qda}. Our results strongly
disfavor a DM candidate annihilating mainly into  $e^+e^-$ with \sigmav $\gsim$ 0.3 (1)$\cdot 10^{-23}$ cm$^3$ s$^{-1}$ 
for masses of about 1-3 (10) TeV. 
A bit looser constraints are obtained for annihilation into muons or tau leptons, the latter roughly at the level which 
is required to fit the AMS-02 data \cite{Yuan:2013eja,Jin:2013nta}. Assuming a relatively large MAGN contribution, like e.g.~in 
the scenario displayed in the bottom-left panel of  Fig.~\ref{fig:upperlimits}, the IGRB would thus allow to firmly rule out a DM 
explanation of the rising positron fraction.
The non-inclusion of the EBL absorption for the BL Lac population might have some effect only in the minimal background scenario. 
From \cite{2010ApJ...712..238F}, one can estimate that the $\gamma$-ray flux at 80 GeV (the highest {Fermi}-LAT energy, see Figs. 1 and 2) 
 for a population at z=0.5(1) is reduced by 20(50)\% from the source to the Earth. The upper limits on  \sigmav would therefore be modified by a negligible 
factor, and only for the highest masses shown in our plots (constrained indeed by the highest experimental data). 
 
As already mentioned, our results improve the limits derived in Ref.~\cite{2012PhRvD..85b3004C} at DM masses above $\sim$100 GeV 
due to the inclusion of ICS; the 'best-fit' scenario displayed in Fig.~\ref{fig:upperlimits} strengthens those limits further by a 
factor of $\sim$2 due to the contribution from MAGN. Our constraints for high masses are significantly stronger than those obtained 
from the ICS contribution alone in an analysis of preliminary {\it Fermi} data \cite{2009NuPhB.821..399C}, and stronger by an  overall 
factor of roughly 4 (for $b\bar{b}$) compared to those obtained for the analysis of 1y IGRB data in a smaller sky region than considered 
here  \cite{2010NuPhB.840..284C}.  
Compared to the {\it Fermi} analysis of possible galactic DM contributions to the IGRB \cite{2012ApJ...761...91A}, with no background
modeling, our conservative limits are comparable for $b\bar{b}$ and $\tau^+\tau^-$ final states (for $m_{\chi} \gsim 20$ GeV). 
Very similar (projected) limits are also found in an independent analysis \cite{Abazajian:2010zb}, which however tends to find slightly 
more stringent constraints on TeV DM models than Ref.~\cite{2012ApJ...761...91A}.
The adopted background modeling in the  {\it Fermi} analysis \cite{2012ApJ...761...91A}, on the other hand, leads to bounds comparable
with our MAGN best fit choice for $m_{\chi} \sim 100$ GeV; for higher masses $m_{\chi} \gsim 1$ TeV, the bounds in the MAGN best fit case 
are a factor of roughly 3 stronger.

Let us finally remark that the bounds on \sigmav we have derived here are consistent with constraints 
from different targets in the sky obtained with very different astrophysical backgrounds and measurement systematics.
In particular, our 'optimistic' limits  on the DM annihilation cross section --  assuming an MAGN flux which is 60\% of 
its maximal theoretical prediction -- are generally competitive with the bounds from the joint Likelihood analysis of 10 dwarf Spheroidal 
galaxies \cite{2011PhRvL.107x1302A}. In fact, for TeV masses and leptonic final states, our constraints are at least comparable 
even under the conservative assumption of a minimal MAGN contribution (though one should note that for such large DM masses, galactic center constraints are
typically even stronger \cite{Abazajian:2011ak}). 
We also obtain comparable results at low masses with bounds on leptonic channels from radio emission \cite{2012JCAP...01..005F}, 
which in turn  depend significantly on astrophysical and cosmological assumptions. At masses $m_\chi \gsim$ 100 GeV our 
bounds from $\gamma$-rays are stronger, independently of the amount of MAGN included in the background. 
Finally, our conservative constraints  are stronger than those obtained from galaxy cluster \cite{2010JCAP...05..025A,2012MNRAS.427.1651H}
 and CMB observations \cite{2011PhRvD..84b7302G}.

 \section{Conclusions} 
 \label{sect:conclusions}
 
In this work, we have considered the contribution of annihilating galactic DM to the IGRB and
presented an updated calculation of the resulting upper bounds on the DM annihilation cross section $\langle\sigma v\rangle$. In deriving these bounds, we took into account both prompt 
and ICS emission for DM-induced $\gamma$-rays. For the astrophysical $\gamma$-ray background, we included all relevant contributions at their minimally expected level, in particular the 
contribution from unresolved MAGN that has been pointed out only very recently. Given that we only consider high galactic latitudes, our results are very robust and barely depend on the 
choice of DM density profile (apart from an overall normalization corresponding to the local DM density, which may either strengthen or weaken those bounds by a factor of at most 3). Let us 
stress that our bounds are very conservative in that they do not include the expected -- and potentially very sizable -- effect from DM substructures at galactic or extragalactic distances.

The resulting IGRB bounds on $\langle\sigma v\rangle$ are still competitive with, or more stringent than,  limits from radio data, CMB or galaxy cluster observations. In particular, our results -- 
along with stringent limits from antiproton data \cite{2009PhRvL.102g1301D,Cirelli:2013hv,PhysRevD.85.123511} --  strongly disfavor a DM-based explanation of the unexpected rise in the 
cosmic ray positron fraction at high energies  that was recently confirmed by AMS-02 \cite{Aguilar:2013qda}, at least if MAGN do not contribute at the most minimal level currently allowed. The 
main reason for this is the sizable ICS emission in $\gamma$-rays that is induced by the high annihilation rate into leptonic channels required to fit the positron data. 
 
 Even more importantly, we have demonstrated that there is significant room for future improvement of limits on the DM annihilation rate derived from the IGRG. In fact, under the optimistic 
 (though by no means unrealistic) assumption that the astrophysical processes considered here contribute more or less at the maximal level consistent with our present understanding, a 
 'thermal' value of $3\cdot 10^{-26}$ cm$^3$ s$^{-1}$ for the annihilation cross section could be excluded up to DM masses of several TeV. Notably, such a limit will be impossible to reach 
 even for far-future ground-based observations of point- or extended sources like dwarf galaxies or the galactic center. Even for space-based observations of dwarf galaxies, which presently 
 sets the most robust and stringent bounds of this kind, it will be extremely challenging for extended future missions to probe thermally produced DM for such large masses.
 
The most important contribution to the IGRB originates from unresolved MAGN, which we have included here for the first time in the context of putting limits on DM annihilation. In order to
 actually exploit the great potential of the IGRB discussed above, a reduction of 
the relatively large theoretical uncertainty on the diffuse $\gamma$-ray flux from unresolved MAGN is thus mandatory. In particular, one would need several tens of detected MAGN in  
$\gamma$-rays to be able to directly determine their $\gamma$-ray luminosity function (which is presently  derived from radio observations); this goal is well in reach for  {\it Fermi}-LAT 
and next generation 
$\gamma$-ray telescopes. At some point, also an improved understanding of the other astrophysical contributions will help to push limits further down -- especially if one can develop the 
underlying modeling to a level that allows to predict those contributions  not only as an all-sky average but as a function of the viewing angle.
Given the expected significantly increased statistics for future  IGRB observations, finally, chances are good that the IGRB will turn out to be one of the most efficient future means of 
probing thermally produced DM.

 
\acknowledgments
We warmly thank A. Bottino, N.Fornengo and L. Latronico  for useful discussions and  
a careful reading of the manuscript.  We also thank J. Siegal-Gaskins for clarifying discussions. 
T.B. and F.C. gratefully acknowledge support from the German Research Foundation (DFG) through Emmy Noether grant BR 3954/1-1.

\bibliography{paper}

\end{document}